\documentclass[twocolumn,prd,aps]{revtex4}

\begin{document}

\title{Casimir Effect for Gauge Scalars: The Kalb-Ramond Case}

\author{F. A. Barone\footnote{e-mail: fbarone@cbpf.br}, L. M. De Moraes\footnote{e-mail: moraes@cbpf.br} and J. A. Helay\"el-Neto\footnote{e-mail: helayel@cbpf.br}}
\affiliation{Centro Brasileiro de Pesquisas F\'\i sicas, Rua Dr.\ Xavier Sigaud 150, Urca 22290-180, Rio de Janeiro, RJ, Brazil}

\date{\today}

\begin{abstract}
	In this work we calculate the functional generator of the Green's functions of the Kalb-Ramond field in $3+1$ dimensions. We also calculate the functional generator, and corresponding Casimir energy, of the same field when it is  submitted to boundary conditions on two parallel planes. The boundary conditions we consider can be interpreted as a kind of conducting planes for the field in compearing with the Maxwell case.	We compare our result with the standard ones for the scalar and Maxwell fields.
\end{abstract}

\maketitle

\section{Introduction}

	When it was proposed, in 1948, the Casimir effect referred only to the atraction between two metalic plates, not charged and placed paralelly each other \cite{Casimir48}. The reason for this attraction was explained as a change of the vacuum energy of the field due to the presence of the metalic planes which modifies the electromagnetic vacuum field modes. Ever since, there has been a huge literature investigating modifications of the vacuum energy of quantized fields due to the presence of boundary conditions on several fields along surfaces \cite{Milton,Miloni,Plunien,Mostepanenko}. Nowadays, Casimir effect is considered as a modification in the vacuum energy of a given quantum field due to the imposition of boundary conditions on this field on one or more surfaces.
	
	In this scenario, there raise many questions concerning the Casimir effect and, paralelly, TCQ with boundary conditions. We would like to mention, for instance, the behaviour of the Casimir energy with the intrinsic features of the field (dependence with spin, mass, etc) and which kind of boundary conditions can be imposed on a given field. 
		
	Motivated by these issues, in this paper we briefly review the quantum theory and we study the correspoding Casimir energy of the rank-2 skew-symmetric tensor field, or commonly, the Kalb-Ramond field \cite{KRref}, submitted to specific boundary conditions. We consider a situation of parallel planes in order to compare our results with standard situations presented in the literature, specificaly, the Casimir configuration (electromagnetic field between conducting planes) and the case of a scalar field with Dirichlet boundary conditions on parallel planes. The last situation is more interesting once both the 2-form Kalb-Ramond and real scalar fields describe a spinless particle, carrying only one on-shell degree of freedom. Comparing these situations, we must have some insight on the influence of possible symmetries on the Casimir effect.  
	
	In addition, despite the Kalb-Ramond field describes a spinless particle, which could be also described, very simply, by means of a scalar field, it appears naturally in strings theory (as a gauge field) and in supergravity (as auxiliary fields), and the task of describing its dynamics, mainly in the quantum context, is not trival. The 2-form gauge field is not commonly studied in the context of second quantization, even without boundary conditions.
		
	In this paper, using standard Fadeev-Popov methods for gauge fields, we shall calculate the generating functional of the Green's functions for the Kalb-Ramond field in $3+1$ dimensions without boundary conditions. We shall also calculate the same generating functional and the Casimir energy for the same field submitted to boundary conditions on two parallel planes, which can be interpreted, in some sence, as a kind of \lq\lq conducting\rq\rq\ plate to the case of Kalb-Ramond, due to its similarity with the case of Maxwell field.
	
	The paper is organised as follows: in Section 2, we establish the quantum theory of the free Kalb-Ramond field without boundary conditions; in Section 3, the quantum theory for the same field submitted to specific boundary conditions on two parallel planes is contemplated; in Section 4, we present the calculation of the Casimir energy per unity of area for the considered boundary conditions previously presented. Section 5 is devoted to our Concluding Remarks

\section{Free Popagator}

	Using the conventions: $\eta^{\mu\nu}=(1,-1,-1,-1)$ and $\epsilon^{0123}=-\epsilon_{0123}=1$, the system we shall study is described by the Lagrangian
\begin{equation}
\label{deflagrangiana}
{\cal L}=\frac{1}{3!}G_{\mu\nu\gamma}G^{\mu\nu\gamma}\ ,
\end{equation}
where
\begin{equation}
\label{defG}
G^{\mu\nu\gamma}=\partial^{\mu}H^{\nu\gamma}+\partial^{\nu}H^{\gamma\mu}+\partial^{\gamma}H^{\mu\nu}
\end{equation}
is the field strength for the Kalb-Ramond field, $H^{\mu\nu}$. It is worthy emphasizing that $H$ is anti-simetric, it is, $H^{\mu\nu}=-H^{\nu\mu}$, and that the Lagrangian (\ref{deflagrangiana}) exhibits the gauge invariance
\begin{equation}
\label{zxc1}
H^{\mu\nu}(x)\rightarrow {H'}^{\mu\nu}(x)=H^{\mu\nu}(x)+\partial^{\mu}\xi^{\nu}(x)-\partial^{\nu}\xi^{\mu}(x)\ ,
\end{equation}
where $\xi^{\mu}$ is an arbitrary vector field\footnote{The $\xi$ field also exhibits an extra gauge invariance which shall be of no relevance for the presente results.}.

	In order to set up the quantum theory for the Lagrangian (\ref{deflagrangiana}), we shall calculate the generating functional of the Green's functions of the theory using standard Faddeev-Popov methods do handle divergent contributions which come from the gauge freedom (\ref{zxc1}).
	
	Choosing a gauge where
\begin{equation}
\label{zxcv1}
\partial_{\mu}{H}^{\mu\nu}(x)-f^{\nu}(x)=0\ ,
\end{equation}
with $f^{\mu}(x)$ being an arbitrary space time function, it can be shown that the Faddeev-Popov deteminant does not depend on the field $H$. Therefore, the development of the Faddeev-Popov method for (\ref{deflagrangiana}) proceeds analogously to the electromagnetic case, and the generating functional for the Kalb-Ramond field reads
\begin{eqnarray}
\label{zxcv3}
W[J]&=&\int{\cal D}H\int{\cal D}f\ \delta\left[\partial_{\mu}H^{\mu\nu}-f^{\nu}\right]\cr\cr
&\ &\exp\left(-\frac{i}{2\alpha}\int\ d^{4}x f^{\mu}f_{\mu}\right)\cr\cr
&\ &\exp\left(i\int d^{4}x{\cal L}+J_{\mu\nu}H^{\mu\nu}\right)\ ,
\end{eqnarray}
where $\alpha$ is a gauge parameter.

	Integrating over $f$ in (\ref{zxcv3}), we arrive at
\begin{equation}
\label{zxc3}
W[J]=\int{\cal D}H\exp\left(i\int d^{4}x\ {\cal L}+{\cal L}_{\alpha}+J^{\mu\nu}H_{\mu\nu}\right)\ ,
\end{equation}
with the gauge Lagrangian, ${\cal L}_{\alpha}$, given by
\begin{equation}
\label{defLalpha}
{\cal L}_{\alpha}=-\frac{1}{2\alpha}\bigl(\partial_{\nu}H^{\nu\mu}\bigr)\bigl(\partial_{\lambda}H^{\lambda}_{\ \mu}\bigr)\ .
\end{equation}

	Using (\ref{deflagrangiana}) and (\ref{defLalpha}), integrating by parts and considering the anti-symmetry of $H$ and $J$, we recast equation (\ref{zxc3}) in the form
\begin{eqnarray}
\label{funcionalKR}
W[J]=\int{\cal D}H\ \exp\Biggl(i\int d^{4}x\ J^{\mu\nu}(x)H_{\mu\nu}(x)+\cr\cr
\frac{i}{2}\int\!\!\int d^{4}x\ d^{4}y\ H_{\mu\nu}(y)K^{\mu\nu,\lambda\rho}(y,x)H_{\lambda\rho}(x)\Biggr)\ ,
\end{eqnarray}
where we have defined the operator
\begin{eqnarray}
\label{defK}
K^{\mu\nu,\lambda\rho}(x,y)=-\delta^{4}(x-y)\Biggl[\left(\eta^{[\mu\bigl[\lambda}\eta^{\nu]\rho\bigr]}\partial^{\kappa}\partial_{\kappa}\right)\cr\cr
+2\left(1+\frac{1}{2\alpha}\right)\left(\eta^{[\nu\bigl[\lambda}\eta^{\kappa\rho\bigr]}\partial^{\mu]}\partial^{\kappa}\right)\Biggr]\ ,
\end{eqnarray}
taking into account it is anti-symmetric by the exchanges $\mu-\nu$ or $\lambda-\rho$.

	The integral (\ref{funcionalKR}) yields
\begin{eqnarray}
\label{funcionalKR2}
W[J]&=&N\exp\biggl(-\frac{i}{2}\int\int d^{4}xd^{4}y\ J_{\mu\nu}(y)\cr\cr
&\ &\times D^{\mu\nu,\lambda\rho}(x,y)J_{\lambda\rho}(x)\biggr)\ ,
\end{eqnarray}
where $D^{\mu\nu,\lambda\rho}(x,y)$ is the Kalb-Ramond propagator whose Fourier transform is found to be
\begin{eqnarray}
\label{defpropagador}
{\tilde D}^{\mu\nu,\lambda\rho}(p)&=&\int\frac{d^{4}p}{(2\pi)^{4}}D^{\mu\nu,\lambda\rho}(x,y)e^{-ip(x-y)}\cr\cr
&=&\frac{1}{p^{2}}\left(\eta^{\mu[\lambda}\eta^{\nu\rho]}+2(1+2\alpha)\ \eta^{\bigl[\mu[\rho}\ \frac{p^{\nu\bigr]}p^{\lambda]}}{p^{2}}\right)\ .\cr
&\ &
\end{eqnarray}
$D^{\mu\nu,\lambda\rho}(x,y)$ is the inverse of $K^{\mu\nu,\lambda\rho}(x,y)$ in equation (\ref{defK}) in the sense that
\begin{equation}
\int d^{4}z  D^{\mu\nu,\alpha\beta}(x,z)K_{\alpha\beta}^{\ \ \lambda\rho}(z,y)=\eta^{[\mu\bigl[\lambda}\eta^{\nu]\rho\bigr]}\delta^{4}(x-y)\ .
\end{equation}

	The result (\ref{defpropagador}) agrees with the ones presented in the literature \cite{Townsend,Barcelos} and obtained by other methods (adjusting appropriately the $\alpha$ parameter).

	With equations (\ref{funcionalKR2}) and (\ref{defpropagador}), the quantum field theory for the free Kalb-Ramond field is completely stablished.

\section{The Kalb-Ramond field with \lq\lq Casimir\rq\rq\ plates}
	
	To impose a boundary condition on the Kalb-Ramond field, let us consider, first, the case of Maxwell field where a perfectly conducting surface $S$ imposes the boundary condition
\begin{equation}
\label{bcM}
n_{\mu}{\tilde F}^{\mu\nu}(x){\bigl|}_{S}=0\ ,
\end{equation}
on the electromgnetic field, with ${\tilde F}^{\mu\nu}$ being the dual of the field strength, ${\tilde F}^{\mu\nu}=(1/2)\varepsilon^{\mu\nu\lambda\rho}\partial_{\lambda}A_{\rho}$.

	Inspired by the boundary condition (\ref{bcM}) for the electromagnetic case, let us consider a kind of \lq\lq conducting\rq\rq\ surface $S$ for the Kalb-Ramond field, by imposing the condition
\begin{equation}
\label{defcondcont}
n_{\mu}{\tilde G}^{\mu}(x){\big|}_{S}=0\Rightarrow n_{\mu}\epsilon^{\mu\nu\alpha\beta}\partial_{\nu}H_{\alpha\beta}(x){\big|}_{S}=0\ ,
\end{equation}
where $n^{\mu}$ is the normal four-vector to the surface $S$, and where we have used that
\begin{eqnarray}
\label{deftildeG}
{\tilde G}^{\mu}(x)&=&{1\over 3!}\epsilon^{\mu\nu\alpha\beta}G_{\nu\alpha\beta}(x)\  \Rightarrow\cr\cr
{\tilde G}^{\mu}(x)&=&{1\over 2}\epsilon^{\mu\nu\alpha\beta}\partial_{\nu}H_{\alpha\beta}(x)\ ,
\end{eqnarray}
which is the dual of the field strength (\ref{defG}).
	
	Now, let us consider the propagator of the Kalb-Ramond field in $3+1$ dimensions submitted to the boundary conditions (\ref{defcondcont}) on two parallel planes located, in our coordinates system, at $z=0$ and $z=a$. It is,
\begin{equation}
\label{defcondCas}
n_{\mu}\epsilon^{\mu\nu\alpha\beta}\partial_{\nu}H_{\alpha\beta}(x){\big|}_{S_{k}}=0\ ,\ k=1,2\ ,
\end{equation}
where $S_{1}$ and $S_{2}$ stand, respectively, for the planes $z=0$ and $z=a$, which are the surfaces where the field satisfies the boundary condition (\ref{defcondcont}), and $n^{\mu}=(0,0,0,1)$ is the four-vector normal to these planes.

	By comparing with the case of the electromagnetic field, the boundary condition we consider for the Kalb-Ramond field can be interpreted, in some sense, as being the one imposed by a kind of \lq\lq conducting plates\rq\rq, analogouslly to the Casimir configuration for the Maxwell field.

	The generating functional of the Green's functions for the KR field submitted to the condition (\ref{defcondCas}) is given by
\begin{equation}
\label{funcional1}
W_{c}[J]=\int{\cal D}{H}_{(c)}\exp\Bigl({\cal L}(x)\Bigr)\ ,
\end{equation}
where ${\cal D}H_{(c)}$ implies that the functional integration is calculated only over field configurations that satisfy the condition (\ref{defcondCas}). Following a procedure developed to calculate functional integrals of the electromagnetic field with boundary conditions \cite{Bordag}, we rewrite the integral (\ref{funcional1}) in the form
\begin{eqnarray}
\label{funcional2}
W_{c}[J]&=&\int{\cal D}{H}{\prod_{k=1,2}}\delta\left[n_{\mu}\epsilon^{\mu\nu\alpha\beta}\partial_{\nu}H_{\alpha\beta}(x){\big|}_{S_{k}}\right]\cr\cr
&\ &\hspace{1.5cm}\times\exp\Bigl({\cal L}(x)\Bigr),\ k=1,2\ ,
\end{eqnarray}
where, now, the integral is taken over all field configurations, and the delta functional,\break $\delta\left[n_{\mu}\epsilon^{\mu\nu\alpha\beta}\partial_{\nu}H_{\alpha\beta}(x){\big|}_{S_{k}}\right]$, kills off the contributions to the integral which comes from field configurations that do not satisfy the conditions (\ref{defcondCas}). 

	The delta functional presented in (\ref{funcional2}) has the Fourrier representation
\begin{eqnarray}
\label{Fourrierdelta}
&\ &\delta\Bigl[n_{\mu}\epsilon^{\mu\nu\alpha\beta}\partial_{\nu}H_{\alpha\beta}(x){\big|}_{S_{k}}\Bigr]=\int{\cal D}B\cr\cr
&\ &\exp{\left[i\int dS_{k}(x_{\perp})B^{k}(x_{\perp})\left(n_{\mu}\epsilon^{\mu\nu\alpha\beta}\partial_{\nu}H_{\alpha\beta}(x){\big|}_{S_{k}}\right)\right]}\cr\cr
&\ &\hspace{2.cm}\mbox{(no sum over k)}\ ,
\end{eqnarray}
where $x_{\perp}=(x^{0},x^{1},x^{2})$, $B^{1}(x_{\perp})$ and $B^{2}(x_{\perp})$ are auxiliary fields of scalar nature whose domains are, respectively, the planes $S_{1}$ and $S_{2}$, and $dS_{1}(x_{\perp})=d^{4}x\delta(x^{3})$, $dS_{2}(x_{\perp})=d^{4}x\delta(x^{3}-a)$, respectively, indicate that we are integrating along the planes $S_{1}$ and $S_{2}$. Replacing the expression (\ref{Fourrierdelta}) in (\ref{funcional2}), we have that
\begin{eqnarray}
\label{funcional3}
W_{c}[J]=\int\int{\cal D}{H}\ {\cal D}B\exp\Bigl({\cal L}(x)\Bigr)\times\hspace{2.5cm}\cr\cr
\exp{\left[i\int dS_{k}(x_{\perp})B^{k}(x_{\perp})\left(n_{\mu}\epsilon^{\mu\nu\alpha\beta}\partial_{\nu}H_{\alpha\beta}(x){\big|}_{S_{k}}\right)\right]}\ ,\ \ 
\end{eqnarray}
where, now, one considers a summation over the indices $k$.

	In order to write the integral above in a more convenient form, we perform the following translation on the $H$ field\footnote{Translations in the $H$ field do not alter the integral (\ref{funcional3}).}
\begin{eqnarray}
H^{\mu\nu}(x)\rightarrow H^{\mu\nu}(x)-\hspace{4.cm}\cr\cr
\int dS_{\ell}(y_{\perp})B^{\ell}(y_{\perp})\left(n_{\mu}\epsilon^{\mu\nu\alpha\beta}\ {\partial\over\partial y^{\nu}}D^{\mu\nu,}_{\ \ \ \alpha\beta}(x,y)\right)\ ,
\end{eqnarray}
where $D^{\mu\nu,\alpha\beta}(x,y)$ is the propagator (\ref{defpropagador}) of the Kalb-Ramond field without boundary conditions. With this procedure, equation (\ref{funcional3}) is written as a product of two Gaussian integrals
\begin{equation}
\label{funcional4}
W_{c}[J]=W[J]\ {\bar W}[J]\ ,
\end{equation}
where $W[J]$ is the functional for the field without boundary conditions, defined in (\ref{funcionalKR}), and
\begin{eqnarray}
\label{funcionalcorrecao}
{\bar W}[J]=\int{\cal D}B\ \exp\Biggl[i\int dS_{\ell}(y_{\perp})\ B^{\ell}(y_{\perp})\cr\cr
\Biggl(-\int d^{4}x\ J^{\alpha\beta}(x)\frac{1}{2}n_{\sigma}\epsilon^{\sigma\gamma\mu\nu}\frac{\partial}{\partial x^{\gamma}}D_{\mu\nu,\alpha\beta}(y,x)\Biggr)\Biggr]\cr\cr
\times\exp\Biggl[\frac{1}{2}\int\int dS_{\ell}(y_{\perp})dS_{k}(z_{\perp})\ B^{\ell}(y_{\perp})\cr\cr
\Biggl(
-n_{\sigma}\epsilon^{\sigma\gamma\mu\nu}\frac{\partial}{\partial y^{\gamma}}\ 
n_{\lambda}\epsilon^{\lambda\rho\alpha\beta}\frac{\partial}{\partial z^{\rho}}
D_{\mu\nu,\alpha\beta}(y,z)
\Biggr)\ B^{k}(z_{\perp})\Biggr]
\end{eqnarray}
is a correction due to the presence of the \lq\lq conducting plates\rq\rq.

	Calculating the functional integral (\ref{funcionalcorrecao}), with the aid of (\ref{defpropagador}), and using the expression (\ref{funcionalKR2}), equation(\ref{funcional4}) becomes
\begin{eqnarray}
\label{funcionalfinal}
W_{c}[J]=N\exp\Biggl[-\frac{i}{2}\int\int d^{4}xd^{4}y\ J_{\mu\nu}(x)\cr\cr
\times \Biggl(D^{\mu\nu,\lambda\rho}(x,y)+{\bar D}^{\mu\nu,\lambda\rho}(x,y)\Biggr)J_{\lambda\rho}(y)\Biggr]\ ,
\end{eqnarray}
where $N$ is a normalization constant and
\begin{eqnarray}
\label{corrprop}
{\bar D}^{\mu\nu,\lambda\rho}(x,y)&=&\frac{1}{4}\int\ \frac{d^{3}p_{\perp}}{(2\pi)^{3}}\frac{1}{L}\ \Lambda(p_{\perp},x^{3},y^{3})\ \epsilon^{3\alpha\mu\nu}\epsilon^{3\beta\lambda\rho}\cr\cr
&\times&\Biggl(\frac{p_{\alpha}p_{\beta}}{L^{2}}\Biggr)e^{-p_{\perp}(x_{\perp}-y_{\perp})}\ ,
\end{eqnarray}
\begin{equation}
\label{defL}
L=\sqrt{p_{\perp}^{2}}\ ,
\end{equation}
\begin{eqnarray}
\label{defLambda}
&\ &\Lambda(p_{\perp},x^{3},y^{3})=\frac{1}{2\sin(La)}\times\cr\cr
&\ &\Bigl[e^{-iLa}\Bigl(e^{iL(|x^{3}|+|y^{3}|)}+e^{iL(|x^{3}-a|+|y^{3}-a|)}\Bigr)\cr\cr
&\ &\hspace{0.9cm}-\Bigl(e^{iL(|x^{3}-a|+|y^{3}|)}+e^{iL(|x^{3}|+|y^{3}-a|)}\Bigr)\Bigr]\ .
\end{eqnarray}

	With the expression (\ref{funcionalfinal}), we can interpret the propagator of the Kalb-Ramond field in the presence of \lq\lq conducting\rq\rq\ plates as being given by the free propagator (\ref{defpropagador}) plus the correction (\ref{funcionalfinal}) due to the boundary conditions.
	
	With the functional (\ref{funcionalfinal}), we have established the quantum theory of the Kalb Ramond field in the presence of the \lq\lq conducting plates\rq\rq.

\section{Casimir Energy}

	In order to calculate the Casimir energy for the Kalb-Ramond field with the condition (\ref{defcondCas}), we first consider the $00$-component of the energy-momentum tensor of this field, $T^{00}$, which is given by
\begin{eqnarray}
T^{00}(x)&=&\bigl({\tilde G}^{0}(x)\bigr)^{2}-\frac{1}{2}{\tilde G}^{\mu}(x){\tilde G}_{\mu}(x)\cr\cr
&=&\frac{1}{4}\epsilon^{0\nu\alpha\beta}\epsilon^{0}_{\ \gamma\lambda\rho}\frac{\partial}{\partial x^{\nu}}\frac{\partial}{\partial x_{\gamma}}\Bigl(H_{\alpha\beta}(x)H^{\lambda\rho}(x)\Bigr)\cr\cr
&-&\frac{1}{2}\ \frac{1}{4}\epsilon^{\mu\nu\alpha\beta}\epsilon_{\mu\gamma\lambda\rho}\frac{\partial}{\partial x^{\nu}}\frac{\partial}{\partial x_{\gamma}} \Bigl(H_{\alpha\beta}(x)H^{\lambda\rho}(x)\Bigr)\ ,\cr\cr
&\ &
\end{eqnarray}
where we have used the definition (\ref{deftildeG}). The Casimir energy is given by the space integral of the $T^{00}$, in the vacuum state, over the region between the plates,	
\begin{eqnarray}
\label{asd2}
E&=&\int_{0\leq x^{3}\leq a}d^{3}{\vec x}\ \langle T^{00}(x)\rangle\cr\cr
&=&\int_{0\leq x^{3}\leq a}d^{3}{\vec x}\lim_{y^{3}\rightarrow x^{3}}\Biggl(\frac{i}{4}\epsilon^{0\nu\alpha\beta}\epsilon^{0}_{\ \gamma\lambda\rho}\cr\cr
&\ &\hspace{1.6cm}\times\ \frac{\partial}{\partial x^{\nu}}\frac{\partial}{\partial y_{\gamma}}\langle H_{\alpha\beta}(x)H^{\lambda\rho}(y)\rangle\bigg|_{x_{\perp}=y_{\perp}}\Biggr)\cr\cr
&-&\frac{1}{2}\int_{0\leq x^{3}\leq a}d^{3}{\vec x}\lim_{y^{3}\rightarrow x^{3}}\Biggl(\frac{i}{4}\epsilon^{\mu\nu\alpha\beta}\epsilon_{\mu\gamma\lambda\rho}\cr\cr
&\ &\hspace{1.cm}\times\ \frac{\partial}{\partial x^{\nu}}\frac{\partial}{\partial y_{\gamma}}\langle H_{\alpha\beta}(x)H^{\lambda\rho}(y)\rangle\bigg|_{x_{\perp}=y_{\perp}}\Biggr)\ ,
\end{eqnarray}
where we used regularization by point splitting in the third spatial coordinate, in addition to the standard regularization in the temporal coordinate considerated implicitly.

	Using equation (\ref{funcionalfinal}), expression (\ref{asd2}) becomes
\begin{eqnarray}
\label{asdfg1}
E&=&\int_{0\leq x^{3}\leq a}d^{3}{\vec x}\lim_{y^{3}\rightarrow x^{3}}\Biggl[\frac{i}{4}\epsilon^{0\nu\alpha\beta}\epsilon^{0}_{\ \gamma\lambda\rho}\cr\cr
&\ &\hspace{1.3cm}\times\ \frac{\partial}{\partial x^{\nu}}\frac{\partial}{\partial y_{\gamma}}\left({\bar D}_{\alpha\beta}^{\ \ \ \lambda\rho}(x,y)\right)\bigg|_{x_{\perp}=y_{\perp}}\Biggr]\cr\cr
&-&\frac{1}{2}\int_{0\leq x^{3}\leq a}d^{3}{\vec x}\lim_{y^{3}\rightarrow x^{3}}\Biggl[\frac{i}{4}\epsilon^{\mu\nu\alpha\beta}\epsilon_{\mu\gamma\lambda\rho}\cr\cr
&\ &\hspace{1.3cm}\times\ \frac{\partial}{\partial x^{\nu}}\frac{\partial}{\partial y_{\gamma}}\left({\bar D}_{\alpha\beta}^{\ \ \ \lambda\rho}(x,y)\right)\bigg|_{x_{\perp}=y_{\perp}}\Biggr]\ ,
\end{eqnarray}
where we discarded a divergent term linear in $a$ and present even without the presence of the plates. This is justified once this term is interpreted as a contribution to the energy which comes from the vacuum without boundary conditions.

	Using equations (\ref{corrprop}), (\ref{defL}) and (\ref{defLambda}), making $y^{3}=x^{3}+\delta$, integrating over $d^{3}{\vec x}$ and taking the limit $\delta\rightarrow0$, we can show that the second term in (\ref{asdfg1}) is a divergent $a$-independent contribution proportional to the plates area $A=\int d^{2}{\vec x_{\perp}}$. Contributions of this kind are interpreted as the self energy of the plates, and can be discarded once they do not produce the Casimir force.

	With these considerations and after a number of manipulations, the first term on the right hand side of (\ref{asdfg1}) gives the Casimir energy per unity of area $A$
\begin{equation}
{\cal E}=\frac{E}{A}=\frac{-i}{2^{3}}\int\frac{dp_{\perp}^{3}}{(2\pi)^{3}}\frac{p_{0}^{2}}{L}\ a\ \frac{e^{iLa}}{e^{iLa}-e^{-iLa}}\ ,
\end{equation}
where the integral above is commonly found in calculations of the Casimir energy for bosonoc fields \cite{Mostepanenko}. The quantity ${\cal E}$ can be calculated performing the Wick rotation,
\begin{equation}
p_{0}\rightarrow ik_{0}\ \,\ \ {\vec p}_{\perp}\rightarrow{\vec k}_{\perp}\ ,
\end{equation}
by defining $L=i\ell$ and using that
\begin{equation}
\frac{e^{-\ell a}}{e^{-\ell a}-e^{\ell a}}=-\sum_{n=1}^{\infty}{e^{-2\ell an}}\ .
\end{equation}
Also, by using spherical coordinates and integrating in the angular variables, we get the Casimir energy per unity of area
\begin{equation}
\label{resultado}
{\cal E}=\frac{\pi}{2^{7}3^{2}5}\frac{1}{a^{3}}\ ,
\end{equation}
which gives the repulsive Casimir force between the planes
\begin{equation}
F=-\frac{\partial{\cal E}}{\partial a}=\frac{\pi}{2^{7}15}\frac{1}{a^{4}}\ .
\end{equation}

	The result (\ref{resultado}), or equivalently (\ref{comparacao}), exhibits interesting features when compared with results for the other well-known bosonic fields: the scalar and electromagnetic ones. To discuss these points, let us denote the Casimir energies for the scalar field for three different configurations: with Dirichlet conditions on the planes, ${\cal E}_{scalar,DD}$, Neumann conditions on the planes, ${\cal E}_{scalar,NN}$, and mixed conditions, ${\cal E}_{scalar,DN}$, (Dirichlet in one plane and Neumann in the other one). Let us also consider the case of electromagnetic field with Casimir configuration, ${\cal E}_{EM,CC}$, (two perfectly conducting parallel plates), the situation with two infinitely permeable parallel plates, ${\cal E}_{EM,PP}$, and the so-called Boyer configuration (one conducting plate and an infinitelly permeable one). Comparing these situations with (\ref{resultado}), we have
\begin{eqnarray}
\label{comparacao}
{\cal E}&=&-\frac{1}{4}{\cal E}_{scalar,DD}=-\frac{1}{4}{\cal E}_{scalar,NN}=\frac{1}{4}\frac{7}{8}{\cal E}_{scalar,DN}\cr\cr
&=&-\frac{1}{8}{\cal E}_{EM,CC}=-\frac{1}{8}{\cal E}_{EM,PP}=\frac{1}{8}\frac{7}{8}{\cal E}_{EM,CP}\ .
\end{eqnarray}

	From (\ref{comparacao}), we can see that, although the Kalb-Ramond field describes a spinless particle, like the Klein-Gordon field, the Casimir energies of these fields have different signs, giving forces in opposite directions, when we take the cases where the scalar field satisfies equal boundary conditions on the planes (${\cal E}_{scalar,DD}$ and ${\cal E}_{scalar,NN}$), similarly to what we have done for the Kalb-Ramond field (\ref{resultado}). Also, (\ref{resultado}) differs in modulus with respect to the scalar cases ${\cal E}_{scalar,DD}$ and ${\cal E}_{scalar,NN}$.

	The scalar field exhibits repulsive force for mixed conditions (${\cal E}_{scalar,DN}$), where we have different boundary conditions on the planes; but, even in this case, the modulus of ${\cal E}_{scalar,DN}$ is different from ${\cal E}$.
	
	The same analysis can be done for the electromagnetc field. The analogous \lq\lq conducting plates\rq\rq\ for the Kalb-Ramond field gives repulsive Casimir force, contrary to the electromagnetic case, where we have atraction for Casimir and two permeable plates configurations. We have a repulsive force for the Boyer configuration which takes different conditions on the plates, contrary to what we have considered for the Kalb-Ramond field.

	The fact that the electromagnetic Casimir energy for two conducting (or permeable) plates is twice the Casimir energy for the scalar field with Dirichlet (or Neumann) conditions on the panes can be interpreted due to the fact that the electromagnetic field has two degrees of freedom and the scalar field just one. In this case, the electromagnetic field is equivalent to two scalar fields, one with Dirichlet and the other with Neumann conditions on the plates. The same analysis does not remain true for the Kalb-Ramond case.

\section{Conclusion}

	In this paper, we have calculated the generating functional of the Green's functions for the Kalb-Ramond field in $3+1$ dimensions in the case it is not submitted to boundary conditions. We have used standard Fadeev-Popov methods and our results agree with the ones presented in the literature and calculated by other methods.
	
	We have also calulated the generating functional of the Green's functions for the Kalb-Ramond field submitted to the conditions of \lq\lq perfectly conducting plates\rq\rq, establishing the quantum theory for the Kalb-Ramond field with these conditions. We have used the previous result to obtain the Casimir energy for the field submitted to the referred boundary conditions and we have found the interesting result that the Casimir energy per unit of area (\ref{resultado}), in this case, is repulsive and lower (in modulus) when compared with the case of the scalar field with Dirichlet or Neumann conditions on the planes. This happens in spite of the fact that the Kalb-Ramond field describes a spinless particle, as the Klein-Gordon field does.

	We would like to poit out that, with the functional (\ref{funcionalfinal}), we could calculate any quantum quantity for the Kalb-Ramond field with the conditions (\ref{defcondCas}).

	It would be interesting to calculate the Casimir energy for the Kalb-Ramond field submitted to conditions analogous the the ones imposed on the Maxwell filed by the presence of permeable plates, and also consider the analogous of the Boyer configuration (Mixed plates).

\section*{Ackowledgements}

	F.A. Barone would like to thank FAPERJ for the financial support.



\end{document}